\documentclass{article}
\usepackage{graphicx}
\usepackage{pict2e}
\usepackage{mathptmx}
\usepackage{hyperref}
\usepackage{color}

\usepackage{float}
\usepackage{authblk}
\usepackage{titlesec}
\date{}
\titleformat{\section}{\large\bfseries}{\thesection}{2em}{}
\titleformat{\subsection}{\large\bfseries}{\thesubsection}{4em}{}
\newenvironment{acknowledgements}{\paragraph{Acknowledgements}}{}
\renewenvironment{abstract}{\paragraph{}}{}
\newenvironment{keywords}{\paragraph{Keywords}}{}
\renewcommand{\and}{$\cdot$}
\newcommand{\beginsupplement}{%
        \setcounter{table}{0}
        \renewcommand{\thetable}{S\arabic{table}}%
        \setcounter{figure}{0}
        \renewcommand{\thefigure}{S\arabic{figure}}%
     }

\begin{document}

\title{\textbf{Modeling the emergence of modular leadership hierarchy during the collective motion of herds made of harems
}}

\author[1,2]{K. Ozog\'any \thanks{Email: ozogany@hal.elte.hu}}
\author[1,2]{T. Vicsek \thanks{Email: vicsek@hal.elte.hu}}
\affil[1]{\normalsize{Department of Biological Physics, E\"otv\"os University, P\'azm\'any P\'eter s. 1/A, 1117 Budapest, Hungary}}
\affil[2]{MTA-ELTE Statistical and Biological Physics Research Group, P\'azm\'any P\'eter s. 1/A, 1117 Budapest, Hungary}

\begingroup
\let\center\flushleft
\let\endcenter\endflushleft
\maketitle
\endgroup

\begin{abstract}
\textbf{Gregarious animals need to make collective decisions in order to keep their cohesiveness. Several species of them live in multilevel societies, and form herds composed of smaller communities. We present a model for the development of a leadership hierarchy in a herd consisting of loosely connected sub-groups (e.g. harems) by combining self organization and social dynamics. It starts from unfamiliar individuals without relationships and reproduces the emergence of a hierarchical and modular leadership network that promotes an effective spreading of the decisions from more capable individuals to the others, and thus gives rise to a beneficial collective decision. Our results stemming from the model are in a good agreement with our observations of a Przewalski horse herd (Hortob\'agy, Hungary). We find that the harem-leader to harem-member ratio observed in Przewalski horses corresponds to an optimal network in this approach regarding common success, and that the observed and modeled harem size distributions are close to a lognormal.}
\end{abstract}

\section{Introduction}
\label{intro}

Like in human communities, several unique species of gregarious animals have developed social structures based on multiple levels of hierarchical organization \cite{grueter,grueter2}. Small groups of closely related individuals can unite in clans which can form bands or loose aggregations. This phenomenon appears in several different taxonomical orders, common examples range from primates \cite{kummer,abegglen}, through elephants \cite{wittemyer} and whales \cite{whitehead,baird} to equids \cite{rubenstein,feh}. The smallest stable sub-unit where strong bonds exist between members can be a family group based on kinship. One basic unit form is a matrilineal family group consisting of one matriarch and her descendants (african elephant \cite{wittemyer}, sperm whale\cite{whitehead}, killer whale \cite{baird}). Another basic form is a one-male reproductive unit, a harem that consists of several breeding females, their subadult descendants, and is dominated and guarded by only one male (Przewalski horses \cite{boyd}, plains zebras \cite{rubenstein}), or it sometimes includes several non-dominant males as well (hamadryas baboons \cite{kummer}, geladas \cite{dunbar}). These highly social animals build stable, sometimes lifetime long communities, based on a complex relationship network and a strong hierarchical order. When a group of animals moves together or makes a collective decision a consistent leadership hierarchy may be observed in many cases, that can serve for a facile flow of information, as it was demonstrated in pigeons \cite{nagy}. The leadership hierarchy in a group may be completely independent from dominance \cite{nagy2}. However, in social animals dominance can be a determining factor in leadership. This is the case e.g. in chacma baboon \cite{king2008}. Regarding the common success the whole group would gain if led by the one with the best knowledge about the good direction to food resources, but this type of leadership was observed only in few cases, e.g. in bottlenose dolphins \cite{lusseau}. Some other traits are empirically shown to affect the individual's chance of becoming a leader, such as central position in the social network \cite{sueur1}, increased nutrient requirements \cite{fischhoff,sueur2} or age \cite{wittemyer2}.

The dynamics of leadership in a complex society can be well described in terms of hierarchical networks, where the leader-follower relationships between group members are associated with directed connections between the nodes. This approach is very useful in understanding collective behaviour. A collective decision is often based on copying the groupmates$'$ individual decisions \cite{petit}, and in the network view the spreading of the copied behaviour can be interpreted as an information flow through the directed edges of the network. In principle a network can have a modular structure in addition to being hierarchical. In the past a few models have been proposed to display these features, however, in those cases the edges between the units were not directed \cite{ravasz,guimera}, thus, did not correspond to leader-follower relationships. Our aim is to construct a model which reproduces the emergence of a modular and hierarchical leadership network,  that is similar to the above introduced phenomenon of the "group of groups" in a collective decision-making context. We consider the case when leadership is mediated by social relations and dominance. We assume that the leader-follower connections are more dense inside the sub-units, as the social bonds are also usually more tight between the sub-unit members, and that the dominant individuals are able to affect the decisions (or movement directions) of their groupmates. In the model we try to find simple rules which cause the emergence of smaller sub-units inside a group. First, we suppose that every individual has an upper limit for the number of bonds he is able to maintain due to the cost of sociality. Developing a social bond can require remarkable time, or maintaining a harem of a given number of females is a costly task for a male. This upper limit of possible bonds introduces a typical sub-unit size. Besides the limitation in bond number, there is also a need for intra-unit cohesive forces that give rise to higher connectivity inside the sub-units.

\begin{figure}[bth]
\centering
\includegraphics[width=0.95\textwidth]{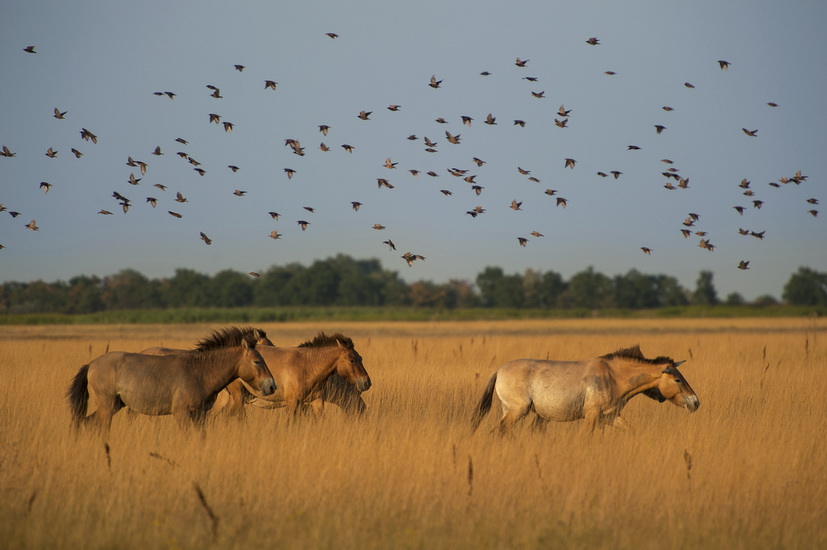}
\caption{The motivation of our model is the understanding of the leadership hierarchy in Przewalski horses. Photo by K. Ozog\'any}
\label{przewalski}
\end{figure}

In large animal groups (bird flocks, fish schools, insect swarms), where individuals can communicate only locally and individual identification is costly, self organizing can be the main driving mechanism of collective decision making processes. In small stable groups, where global communication is possible and complex social relationships can develop between members, sociality is more important in leadership \cite{king2011,jacobs}. Our model on the leadership hierarchy is a combination both of the above mechanisms, since this can be a suitable approach for the case of social animals living in big groups \cite{petit2}. As a result, it leads to the spontaneous emergence of a modular hierarchical network underlying a group composed of sub-groups. The observation of the collective movements of a free-ranged herd of Przewalski horses (Equus ferus przewalskii, Fig.~\ref{przewalski}) helps us in defining the rules of the algorithm. This herd consists of stable and non-overlapping harems, each of them being guarded and herded by one stallion. The observed collective motion pattern of the herd shows the borderlines of the individual harems through the harem members$'$ cohesive motion (Fig.~\ref{horses}). Thus, it can be assumed that the leadership is affected by the horses$'$ social bonds. We aim to reproduce the special case of the leadership hierarchy of a wild horse herd. Some aspects of the behaviour of harem-living animals are used to formulate social rules, thus our model is likely to be applicable for other species as well.

\section{Observations}
\label{observation}

It was found in plains zebras that movement initiations inside the harems are determined by a consistent hierarchy of the individuals, and the position during travelling correlates with the initiation order \cite{fischhoff}. In addition, it was shown that movements of the herd are dominated by lactating females at two different levels of social organization. The direction of motion of individual harems were initiated by lactating females, while the motion of the herd was likely to be determined by harems containing more lactating females \cite{fischhoff}. Both of plains zebras and Przewalski horses live in similar social organization, in a fission-fusion system of harems and bachelor groups. Hence, we suppose that a consistent leadership hierarchy can exist over movements among the harems of a Przewalski horse herd, as well. In order to get an insight into this phenomenon we make an estimation based on aerial images, using the assumption that the harems$'$ relative position occupied in the herd while moving can be an appropriate indicator of the rank in the leadership hierarchy.

\begin{figure}[tb]
\centering
\includegraphics[scale=0.6]{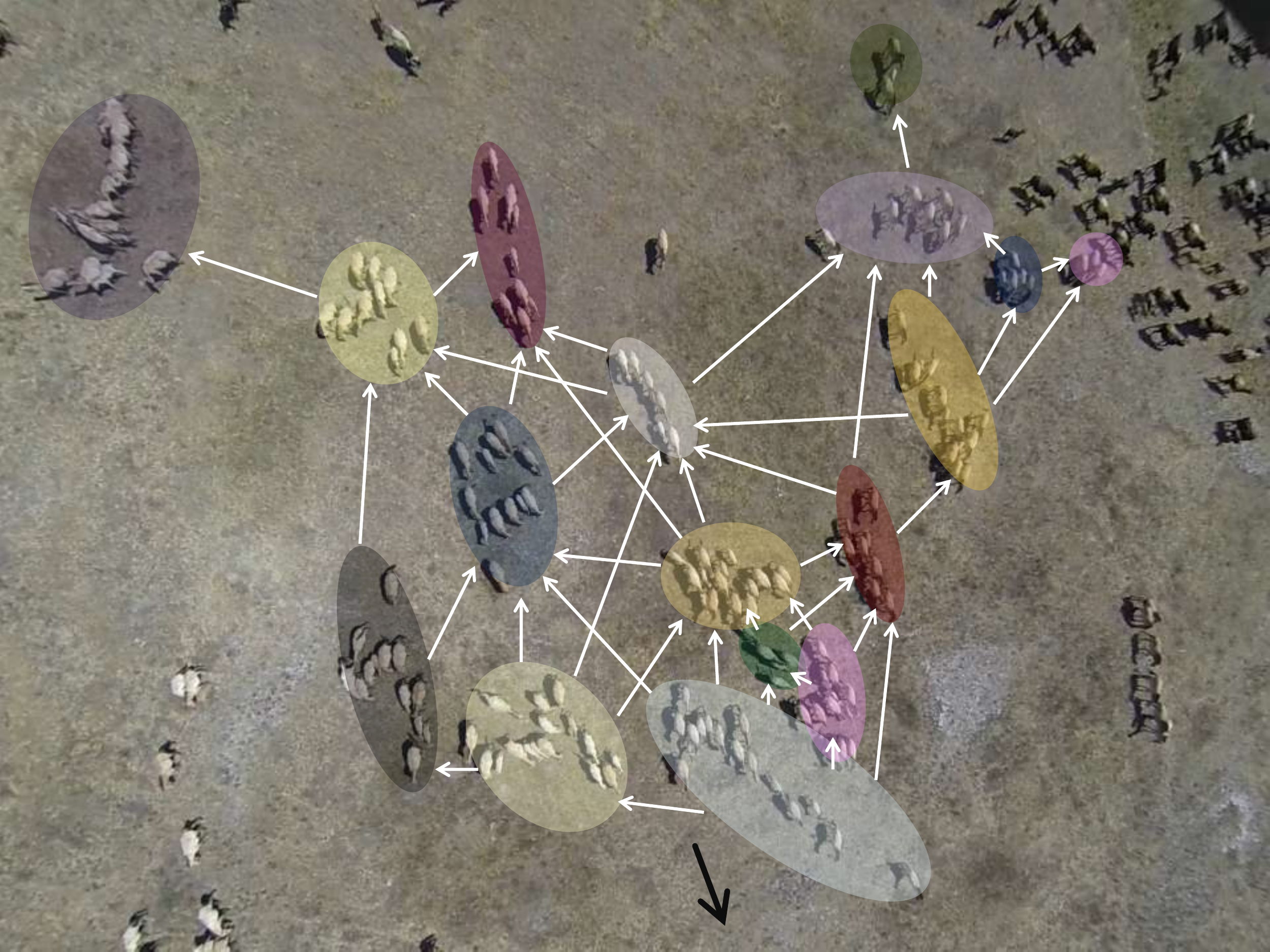}
\caption{Przewalski horse herd during movement ($n\approx150$). The spatial distribution of the herd roughly shows the borderlines of the harems. Colored areas emphasize the individual harems identified due to the harem members$'$ cohesive motion. White edges point from the leader to the follower harem, and are defined between the neighbouring harems within a given range. As the herd takes on a V formation, from two harems connected by an edge, the one closer to the tip of the V is identified as leader. If both are a similar distance from the tip then the more centered one is the leader. We base this leadership definition on the observation that in plains zebras the individuals in the front are more likely to lead \cite{fischhoff}. The black arrow shows the direction of motion. The bachelor groups in the bottom left corner and on the right, and the oxen in the top right corner are not considered.}
\label{horses}
\end{figure}

We use small flying robots and a blimp filled by helium to capture aerial records of a cohesive herd of Przewalski horses living under semi-reserve conditions in the Hortob\'agy National Park, Hungary. The herd has around $240$ individuals, and consists of stable and non-overlapping harems, with sizes ranging from $2$ to $19$, including the harem stallion. There are less stable bachelor groups surrounding the herd of harems, that are not considered. The spatial distribution of the horses roughly shows their social organization, since harem members keep closer to each other than the typical distance between the harems (Fig.~\ref{horses}). During the collective movements the harems remain as cohesive units inside the herd and the leadership roles among them seem to hold on for several minutes (\href{http://youtu.be/-L8ddPLojDc}{Movie S1}). At longer time scale the leadership may fluctuate to a certain extent, but it may have a well defined structure on average, similarly to the individual hierarchies reported in other species \cite{nagy,akos}.

The estimated leadership network of the harems is demonstrated in Fig.~\ref{horses}.,  and its hierarchical layout is visualized on the left side of Fig.~\ref{comparison} (a) using the reaching centrality method \cite{mones}. We have analysed several images recorded on different days, and they show a sign of some consistency (Fig.~\ref{s1} and \ref{s2}). Our definitions of the nodes and edges of the harem network are the following. A node represents a harem. Individuals who are within one horse length of each other, or keep together during moving, belong to the same harem. As a verification of our harem definition the number of adult and infant horses in each harem was compared with the catalogue of harems established by the national park. Harems which are within a given interaction range of each other, and are seen directly (i.e. not covered by another harem) by at least one of the pair, are linked with an edge. The envelope of the herd is a V formation, pointed in the direction of the movement. Thus from two harems we define as leader the one which is closer to the tip of the V. In the case when both of them are roughly the same distance from the tip, then the more central one is defined as leader. The direction of an edge points from the leader to the follower harem. This is only a first estimation of the precise leadership hierarchy, which is to be determined through a deliberate analysis of individual tracks. The identification of the harems includes some uncertainty as well, since harem members disperse sometimes for short time intervals making the finding of the borderlines of the harems in a still image difficult.

The right side of Fig.~\ref{comparison} (a) shows two enlarged harems including their estimated internal leadership structure of the individuals. The edges between the individuals are defined in a similar way as the ones of the harem network, based on the individuals$'$ position occupied in the particular harem on the still picture. We denote as the leader of the harem (L) the individual that is closest to the tip of the harem, and denote as follower harem members (F) all the other individuals further from the tip. We define an edge between two individuals if they are seen directly by at least one of the pair, with the direction pointing from the leader individual closer to the front of the harem towards the follower individual.

\section{Model description}
\label{model}

The starting point of our model is the agent-based model of \cite{nepusz}. It consists of individuals facing a problem solving situation iteratively in each time step. They have an environment that changes slowly in time over several discrete states. The individuals aim to guess the actual state of the environment, and the ones that guess correctly gain score. They have diverse abilities to guess the state of the environment and are also allowed to seek advice from other groupmates. In the context of collective motion the environment can represent the seasonally changing habitat of the group, including the actually accessible drinking or feeding places, and a good choice represents a good direction towards it. More generally, the environment can be any decision making situation, where individuals can choose from several discrete decisions. The proper choice is the appropriate behaviour, which enables to benefit from the situation. In order to keep the herd$'$s cohesiveness, the individuals should synchronize their behaviour, which can be achieved by copying groupmates. Every individual in the model is likely to consider the decision of several other groupmates and ponders over them, taking into account its own guess as well. Finally it decides on its own, or copies a groupmate$'$s decision, who seems to have better ability in guessing. Each individual estimates the perceived ability of others from whom it already copied a decision based on the guessing success, and tries to follow those who seem to be the best. As the model proceeds stable leader-follower relationships emerge between the individuals, where the follower individual considers the decision of the leader individual when making his own decision. In other words, the model leads to spontaneous emergence of a leadership hierarchy, where the leadership hierarchy is defined through the information flow: we assign an edge where information exchange is between two individuals, with the arrow directed from the source of information to the target.

In order to arrive at a herd which consists of loosely connected harems, we introduce two types of individuals into the model showing different behaviour. The two types are called leader-type (L) and follower-type (F) individuals. The leader-type individuals have some properties that make them likely to become the leader of a sub-unit. In addition, they have organizational skills, which increase the sub-unit$'$s cohesiveness. In Przewalski horses the stallions play a dominant role by keeping the harem together, but the direction of motion can also be strongly influenced by a dominant mare. For the sake of simplicity, in the model we assign these two particular roles (leading in movements and sub-unit managing role or herding) to one special individual type, the leader-type individual. It is important to note, that the model does not assume that in horses the stallion leads the harem, only the fact that there is one individual in each harem that has more influence in leading than the others, and that there are intra-unit cohesive forces in the harems that give rise to higher connectivity among members. The leader-type to follower-type ratio is preferably unequal. Every individual of both types tries to maximize its own guessing success by following those who seem to be better in finding the good answer. Leader-type individuals, in addition, try to collect followers and build non-overlapping harems of follower-type individuals due to herding. Three herding behaviours are integrated into the model. First, the leader-types do not favour if their harem members follow individuals from another harems. They try to prevent these leader-follower relationships and thus try to prevent their followers to leave their harem. This behaviour serves as a cohesive force inside the harems. The leader-types also try to prevent other leader-types to follow their harem members. Finally, every leader-type individual tries to increase its harem size by herding those follower-type individuals who are not belonging to any harem yet. The two types of individuals differ in their typical ability in guessing the right state of the environment. The largest ability values are matched to leader-type individuals enabling them to assume higher positions within the hierarchy. Costly tasks are incorporated as limitations in the model. The cost of having links with other individuals is included such a way, that every individual has a link capacity, which is an upper bound of outgoing edges they are able to establish. The two types do not differ in their typical link capacities. For the leader-types, its value also determines the fitness of the individual, since the frequency that the individual shows herding behaviour and the probability that it succeeds in herding is proportional to its link capacity. On the other hand, it determines the maximum harem size of a given leader-type individual. Thus the cost of fighting and herding is included in the link capacity as well.

The model starts from naive individuals without relationships, who do not know the abilities of others. As it proceeds the individuals learn about each others$'$ abilities and try to copy decisions from the ones who seem to be more successful in guessing (\href{http://youtu.be/V7ehrKq_6VI}{Movie S2}). The resulting hierarchy of a single run is a realization of a leadership structure between the given individuals, which emerges spontaneously, and develops to a stable and beneficial configuration. It is a model of a natural animal leadership structure achieved during a given time of living together and remaining quite consistent.

\subsection{Formal model description}
\label{formal}

\subsubsection{Individuals and their environment}

The model contains $n$ individuals, of which $m$ are assigned as leader-type and $n-m$ as follower-type. They are embedded in an environment that changes in an unpredictable way, and the individuals have to guess its state to gain benefit. The environment is always in exactly one of the $l$ possible discrete states. The simulation consists of steps, and the state of the environment is constant within a step but may change between steps with probability $p$. Thus, the characteristic time between state-flips is about $1/p$ steps, during which the individuals can "learn" about the environment$'$s actual value. When environment changes state, the given state is replaced uniformly by a randomly chosen other one. 

Within each step, individuals must make a decision in the sense, that they must guess the actual state of the environment. Each individual has a predefined ability (e.g. $a_{i}$ denotes the ability of individual $i$), which is the probability that the individual can make a proper guess of the state of the environment without any external information. If individual $i$ decides on its own (without copying anyone), than it chooses the current state of the environment with probability $a_i$, or chooses any other state uniformly from the non-correct states of the environment with probability $1-a_i$. Each individual has the information about its own ability, but does not know a priori the abilities of other individuals. 

The final decision of the individuals concerning the actual state of the environment in each round is composed of their own guess and from some copied decisions of other individuals. The number of other individuals a given person tries to copy in a step is a predefined value and depends on its own ability. Formally, individual $i$ considers the decision of $q(1-a_i)$ other individuals (rounded up to the nearest integer), where $a_i$ denotes its own ability and $q$ denotes the maximal number of groupmates someone can ask in a round. Individuals with higher abilities will try to copy a smaller number of other individuals, since they confide in their own guess. Each individual has an upper limit of the number of other individuals whom he can provide information in a single round. This upper bound is called link capacity (e.g. $c_{i}$ denotes the link capacity of the individual $i$). 

The individuals have information about the type, but not about the ability and link capacity of others. They can estimate the ability of those whom they have already copied. The estimation of another$'$s ability can be interpreted as a trust in someone$'$s decision. The values of the estimated or perceived ability of others are stored in a matrix, where the element $t_{ij}$ denotes the ability score of individual $j$ as perceived by individual $i$, and is calculated by using the rule of succession 
\begin{equation}
t_{ij} = \frac{s_{ij} + 1}{n_{ij}+l},\label{tij}
\end{equation}
where $s_{ij}$ is the number of rounds in which individual $i$ received a correct answer from individual $j$ and $n_{ij}$ is the number of rounds when individual $i$ received any information from individual $j$. Thus, $t_{ij}$ is an estimation by individual $i$ of success rate of individual $j$ in finding the correct answer. The $t_{ii}$ diagonal element of the perceived ability matrix is equal to $a_i$, since every individual knows its own ability exactly. In each round individual $i$ considers copying those who it perceives as the best (i.e. $q(1-a_i)$ other individuals with the highest $t_{ij}$, $j=1,2,...,n$ perceived ability scores). The perceived ability score also determines the weight of the copied answer in a person$'$s final decision. The leadership hierarchy emerges from the information flow between the individuals, where a directed edge of the hierarchical structure conveys a copied answer from an individual to another. The precise mechanism of individual decisions is described in the next paragraph.

Individuals who can infer properly the current state, gain a positive feedback at the end of the step (they gain $1$ score), the ones who do not infer the state, gain a negative feedback (they gain $0$ score). The total score of an individual is the exponential moving average of all the feedbacks gained during the simulation with a half life of $50$ steps (this means that in the average the weight of a feedback vanishes exponentially with time, e.g. a feedback received $50$ steps ago has a weight of $0.5$).

Both the abilities and the link capacities of the individuals are diverse, and are random variables chosen from a given distribution. The abilities are drawn from a bounded Pareto distribution, since the fat-tailed distributions maximize group performance \cite{zafeiris} and promote the emergence of hierarchy \cite{nepusz}. From the ordered abilities the $m$ largest values are assigned to leader-type individuals and the rest to follower-types. The higher ability values enable the leader-type individuals to assume higher ranking positions in the hierarchy. If the abilities of the leader-type and follower-type individuals would be drawn from different distributions, then when changing their ratio, the overall ability distribution of the group would change. The results would not be comparable to each other, because the ability distribution has a notable effect on the overall success \cite{nepusz}. The link capacities are drawn from several different distributions to test their effect on the sizes of the emerging sub-units, namely uniform, Poisson, delta and lognormal distributions with varying parameters. The short descriptions and the default values of the input parameters used in the simulations are listed in Table~\ref{input}.

\begin{table}[t]
\begin{center}
\caption{Description and default values of the input parameters used in the simulations}
\label{input}
\begin{tabular}{p{2cm} p{5cm} p{7cm}}
\hline\noalign{\smallskip}
\textbf{Notation} & \textbf{Description} & \textbf{Default value}  \\
\noalign{\smallskip}\hline\noalign{\smallskip}
$n$ & the number of all individuals & $n = 200$ \\ \noalign{\smallskip}\hline\noalign{\smallskip}
$m$ & the number of leader-type individuals & $m = 25$ \\ \noalign{\smallskip}\hline\noalign{\smallskip}
$l$ & the number of states in the environment & $l = 5$ \\ \noalign{\smallskip}\hline\noalign{\smallskip}
$p$ & the probability of state change in the environment & $p = 0.1$ \\ \noalign{\smallskip}\hline\noalign{\smallskip}
$\eta$ & noise & $\eta = 0$ \\ \noalign{\smallskip}\hline\noalign{\smallskip}
$a_1,a_2,...,a_n$ & the abilities of the individuals & random variables distributed according to a bounded Pareto with $\mu= 0.25$ average value, $\sigma= 1/\sqrt{48}$ standard deviation and an upper limit of $1$\\ \noalign{\smallskip}\hline\noalign{\smallskip}
$c_1,c_2,...,c_n$ & the link capacities of the individuals & random variables distributed according to a discrete uniform distribution between $(3,20)$ \\ \noalign{\smallskip}\hline\noalign{\smallskip}
$s_1,s_2,...,s_n$ & the fitness of the individuals & $s_i=c_i, i=1,2,...,n$ \\ \noalign{\smallskip}\hline\noalign{\smallskip}
$t_{ij}$ \hspace{1cm} $i,j=1,2,...,n$ & the ability score of individual $j$ perceived by individual $i$ & initially $t_{ij}=1/l$, and is calculated after each round by using (\ref{tij})\\ \noalign{\smallskip}\hline\noalign{\smallskip}
$q$ & the maximal number of individuals that someone can consider for copying & $q=3$ \\ \noalign{\smallskip}\hline\noalign{\smallskip}
$r$ & the herding frequency of leader-type individuals & $r = 0.5$ \\ \noalign{\smallskip}\hline\noalign{\smallskip}
\noalign{\smallskip}
\end{tabular}
\end{center}
\end{table}

\subsubsection{Decision making}

Each single round of the simulation consists of the following phases:
\begin{enumerate}
	\item The state of the environment is calculated: the state of the previous round remains with probability $1-p$, or changes to a different state with probability $p$. Individuals make their own guesses about the state of the environment based on their abilities.
	\item The leader-type individuals do their herding behaviour which is described in the next section. $rn$ herding attempts take place in each round, where $r$ denotes the herding frequency, ranging from $0$ to $1$, and $n$ denotes the number of all individuals.
	\item Every individual nominates $q(1-a_i)$ other individuals (rounded up to the nearest integer) whom it wishes to copy in the current round. Individual $i$ considers the $i$th row of the perceived ability matrix (the $t_{ij}$ elements, $j=1,2,...,n$), and nominates the individuals with the $q(1-a_i)$ highest perceived ability values.
	\item Every individual accepts at most $c_i$ other individuals from those who nominated him, preferring those whom he has already passed information in the previous rounds. Individuals propagate their decision of the environmental state from the previous round to the accepted ones. It is important that the decisions from previous round are passed, because the propagation of decisions from the current round could not be solved in a synchronical way. The remaining (not accepted) individuals who nominated the given person will not get any information in this round from him. Individuals maintain a taboo list, and do not nominate again in the next round the ones who can not propagate an answer to them.
	\item Individuals make their decisions by calculating the majority opinion taking into account the answers of the copied individuals and their own guess. Individual $i$ weights the informations about the environmental state, where the weight of a copied answer from individual $j$ is the $t_{ij}$ perceived ability score, and the weight of its own guess is its $t_{ii}=a_i$ ability. Then it chooses the answer with the highest summarized weight. 
	\item Individuals get a feedback about the current state of the environment, and they re-calculate the perceived abilities of others by updating the $n_{ij}$ and $s_{ij}$ counters in (\ref{tij}).
\end{enumerate}
The above described single round repeats iteratively in the simulation. In each round a directed network is defined between the individuals, where the edges refer to  copied decisions. In other words, if individual $j$ considered the decision of individual $i$ with some weight in his choice than an edge is added to the network with $i$ being the source and $j$ being the target node (i.e. $i$ being the leader and $j$ being the follower). The leader-follower relationships occur spontaneously from the small initial perturbation of the perceived ability matrix and gradually develop to a stable configuration.

A random noise is included into the model. Before nominating other individuals for copying we add Gaussian noise with zero expected value and a given standard deviation to the perceived ability values.

\subsubsection{Leader-type individuals}

Every individual has a predefined type, they are either leader-type (L) or follower-type (F). Individuals know not only their own type, but also the type of the others. The leader-type individuals are responsible for the forming of sub-units or harems. The harems are identified through the leader-type individuals, e.g. the $h_i$ harem denotes the harem led by the $L_i$ leader-type individual. Each individual is aware of which harems it belongs to. An $L_i$ leader-type individual belongs to its own $h_i$ harem. An $F_j$ follower-type individual belongs to the $h_i$ harem, if $F_j$ considered the decision of $L_i$ in the current round (in other words $F_j$ followed $L_i$). Thus, the $h_i$ harem is the union of the $L_i$ leader-type individual and all the follower-type individuals that followed $L_i$ in the current round. These individuals are also called the members of the $h_i$ harem. The size of a harem is the number of individuals belonging to the harem (including the leader as well). Note, that with this definition a follower-type individual can belong to one, more or zero harems. In this latter case it is called a lonely individual. Individuals have a predefined $s_i$ fitness parameter, which in the case of the leader-types, characterizes their potential of maintaining a harem, in the sense that the frequency and success rate of their herding attempts is proportional to their fitness. The fitness of $L_i$ is equal to its link capacity, $s_i=c_i$. Thus, the link capacity parameter determines, on one hand, $L_i$$'$s  success rate in herding, and on the other hand, an upper bound for the possible size of the harem that $L_i$ is able to keep together. A herding attempt consists of the following three steps:
\begin{enumerate}
\item An $L_i$ leader-type individual is chosen randomly for herding with probability proportional to its $s_i$ fitness.
\item	For $L_i$ chosen for herding, a list of follower-type individuals is identified. The list includes those members of harem $h_i$ who also follow an individual from a different $h_k$ harem (either an L or an F individual), or are followed by another $L_m$ leader-type individual. The list includes also all the lonely individuals in the case if the herding $L_i$ has free links. In other words, all the possible edge operations are identified, including the adding of new edges and the deleting of undesired edges.
\item	A possible edge operation is chosen randomly with uniform probability. If an edge is chosen for adding, directed towards a lonely individual, than the herding $L_i$ reinforces its ability score perceived by the $F_j$ lonely individual. Formally, it means that both the $s_{ji}$ and $n_{ji}$ values in (\ref{tij}) are increased by $1$ in the $t_{ji}$ perceived ability score. This has an equivalent effect to $L_i$ passing to $F_j$ a proper answer about the state of the environment in the decision making process described in the previous section. Since after this attempt the ability score of $L_i$ perceived by $F_j$ increases, $F_j$ is more probable to follow $L_i$ in the next rounds. If an edge is chosen for deletion, three cases can occur. In the first case, $F_j$ from the $h_i$ harem follows $F_k$ from a different harem. $L_i$ succeeds in breaking this edge randomly with a probability proportional to its $s_{i}$ fitness. Breaking a relationship between a follower $F_j$ and a leader $F_k$ means to reset the perceived ability score $t_{jk}$ to its initial value. In the second case, the undesired edge points from another $L_k$ to an $F_j$ from the $h_i$ harem (in other words $F_j$ follows another $L_k$ beside $L_i$). This time $L_i$ and $L_k$ $"$fight$"$ for $F_j$, and $L_i$ wins over $L_k$ with probability $s_{i}/(s_{i}+s_{k})$ and looses with probability $s_{k}/(s_{i}+s_{k})$. Then the winner individual reinforces his ability score perceived by $F_j$, and breaks the edge between $F_j$ and the losing individual. In the third case, the undesired edge points from $F_j$ from the $h_i$ harem to another $L_k$ ($F_j$ is followed by $L_k$). $L_i$ and $L_k$ $"$fight$"$ for $F_j$ as in the previous case.
\end{enumerate}

\subsubsection{Group performance and structural properties of the network}

To characterize the overall success of the herd, we quantify the common performance of the group through the number of times when the individuals could infer the actual state of the environment correctly. Each individual in each time step scores $1$ if successfully guessing the environmental state and $0$ if not. The $p_i$ performance of an individual is the exponential moving time average of his score history, with a half life of $50$ steps. This means that the scores of the past rounds are weighted with an exponentially decaying factor, where the score of the current round has a weight of $1$, while the score $50$ steps ago has a weight of $1/2$. Individual performance is thus simply the ratio of proper guesses of someone, where past guesses have less weight as time passes. The performance of the whole society is the average performance of the individuals. Without the possibility of copying the performance of an individual would be his $a_i$ ability on average. According to this, the relative improvement of the overall performance - compared to a group without copying - can be defined as the common performance minus the average ability, divided by the average ability expressed in percentages. Formally the relative performance improvement which we use to measure the success of the society is:
\begin{equation}
P=\frac{\left\langle p_i\right\rangle - \left\langle a_i\right\rangle}{\left\langle a_i\right\rangle}\cdot 100\%\label{imp},
\end{equation}
where $a_i$ are the abilities, and $\left\langle \right\rangle$ denotes averaging over individuals. A relative performance improvement of $0\%$ thus means, that the group is performing at the same level as a group without copying, and a relative performance improvement of $100\%$ means that it performs twice as well.

In order to characterize the extent of the hierarchical organization of the network we use three measures, the fraction of noncyclic edges, global reaching centrality \cite{mones} and the fraction of the largest cycle-free arc set \cite{eades}. The reaching centrality method is based on the assumption that the rank of the nodes is related to their impact on the whole network. A node$'$s impact can be quantified with its local reaching centrality, which is the proportion of all nodes reachable from it via outgoing edges. The global reaching centrality (GRC) of a network is related to the heterogeneity of the local reaching centrality distribution of its nodes which is wider for hierarchical structures. Thus, the quantity GRC measures the level of hierarchy of a network, with a value close to zero corresponding to no hierarchy, while GRC being about $1$ signalling a highly hierarchical network. 

In order to get a qualitative insight of the layout of the emerging leadership network, we visualize the individual and the harem network with the reaching centrality method \cite{mones}. This visualization method is based on the local reaching centralities of the nodes, where nodes with similar values lie in the same layer, and the one with the highest value is on the top. To reveal modular structure and find the communities we use the clique percolation method and the CFinder software for visualization \cite{adamcsek,palla}. The convergence is indicated by the fraction of changed edges in a time step going to zero, and the harem sizes converging to stable values.

\section{Results}

As the model proceeds iteratively, a network of leadership emerges. The individual leadership network consists of nodes representing the individuals, and the directed edges between them show whose decision was copied by whom in previous rounds. Since as a consequence of herding sub-groups are expected to emerge led by a leader-type individual, we define the harem leadership network in the simplest way, by considering only the subgraph of the leader-type individuals. In the harem leadership network the leader-type individuals are the nodes, and the connections between them are the edges. Note that the edges in the empirical network of Fig.~\ref{horses} denote the possible leader-follower relationships, and not the realized ones, by definition. A given harem $i$ has an incoming edge pointing from harem $j$, if $j$ can be seen from and thus followed by $i$, but it does not necessarily mean that $i$ followed $j$. The situation is the same in the model, an individual (or a harem) $i$ has an incoming edge from $j$ if its decision is considered by $i$. But in making its own decision $i$ weights all the considered decisions, and may finally choose a different decision than $j$. 

\begin{figure}[p]
\centering
\includegraphics[width=0.8\textwidth]{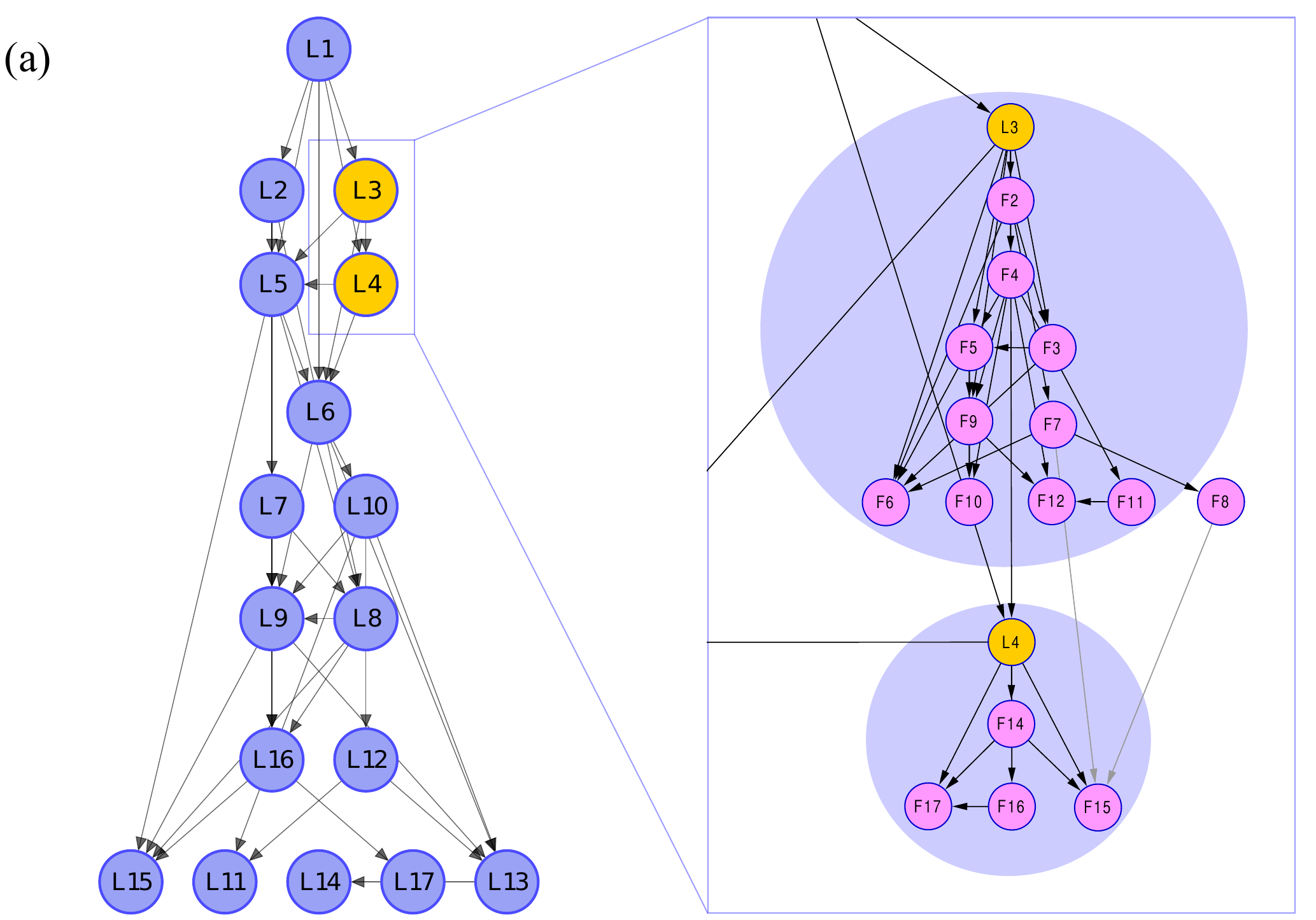}\\\hspace{0.1cm}\includegraphics[width=0.8\textwidth]{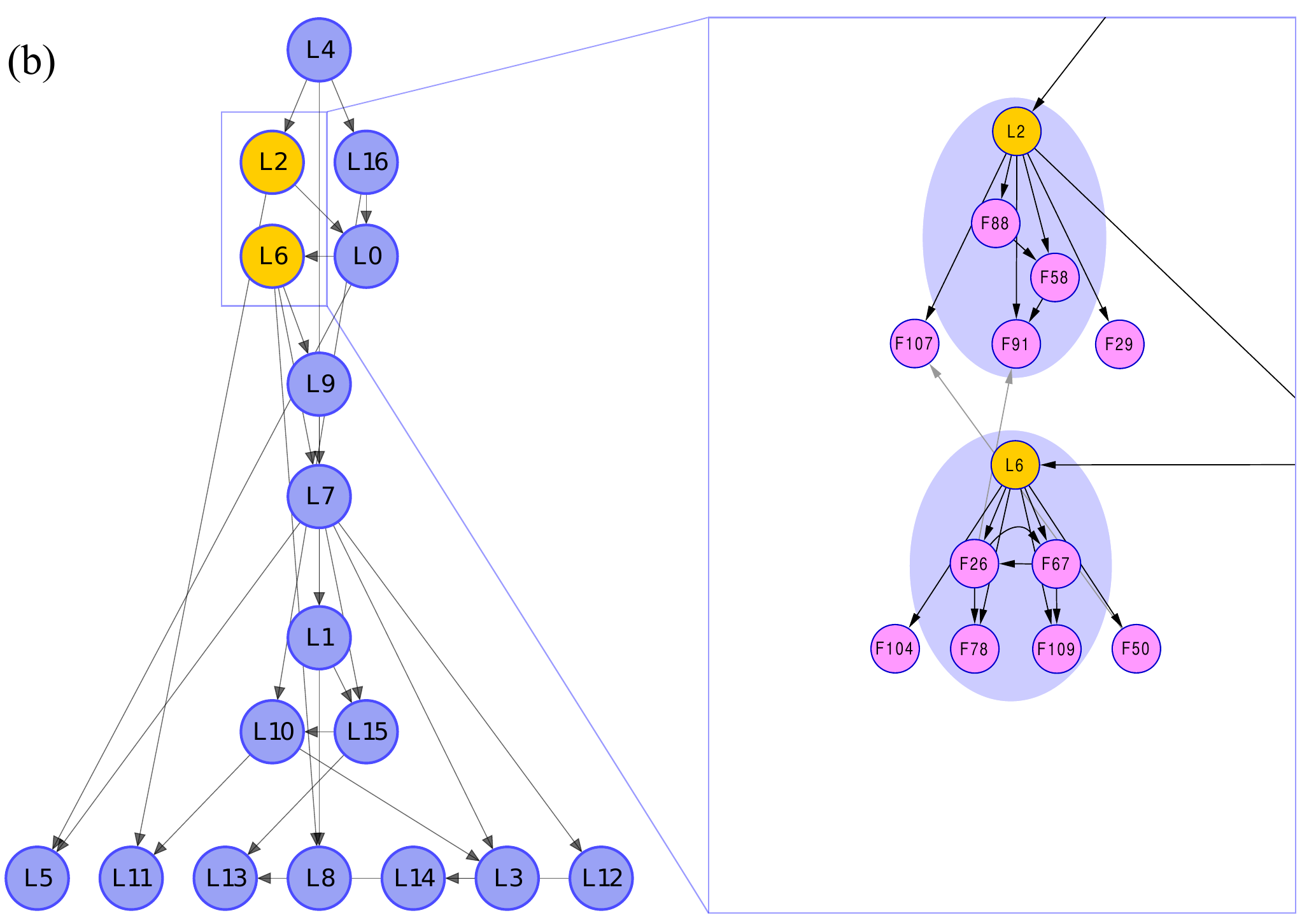}
\caption{A typical leadership hierarchy among harems (a) in the experiment of the Przewalski horses based on Fig.~\ref{horses} and (b) a very similar network resulting from our model. The nodes with names starting with L denote harems of a given harem leader, and the directed edges point from the leader to the follower harem, showing thus the flow of information. Visualization is performed by the reaching centrality layout \cite{mones} with $z = 0.1$. The global reaching centrality values of the two harem networks are similar, 0.65 and 0.67 for the experimental, and for the model, respectively, which indicates a similar extent of hierarchy. The enlargements of the blue square areas reveal the internal connections of individuals inside some harems. The pink nodes with names starting with F denote the harem members, and orange emphasizes the harem leaders whose harem is enlarged. Light blue areas denote communities identified with CFinder \cite{adamcsek}. Relationships between two harem member individuals connecting two different harems are indicated in gray. The number of harems in the simulation is $m=17$, and the number of all individuals is $n=150$, as it is in the experiment, the number of edges between the harems is $26$ and $36$ in the model and in the experiment, respectively.}
\label{comparison}
\end{figure}

One outcome of our model is that the leadership network, both on the level of individuals and on the level of harems, converges to a stable state. Convergence is reached after several hundred time steps and it is indicated by the convergence of the values of the underlying perceived ability matrix, and by the fact that the fraction of changed edges over a time step goes to zero. In the presence of a small amount of noise the resulting network can fluctuate slightly around an average, similarly to a real group where leadership roles can be flexible. The typical structure of the converged networks is qualitatively similar to a herd consisting of harems. Examining them with CFinder \cite{palla,adamcsek}, which uses the clique percolation method, communities can be revealed that are associated with more highly inter-connected subgraphs, typically made of $k=3$ and $k=4$ cliques. Two typical community type occurs in the networks. The first typical community consists of one leader-type and some follower-type individuals. The same leader-type individual is often participating in more of such communities, thus the union of these communities can be viewed as a harem that is led by the leader-type individual. The communities forming a big harem are often overlapping with many shared nodes, most commonly one or two $4$-clique communities are embedded in a $3$-clique community. The harem leaders (besides their harem forming communities) often form a second type of communities that consists only of leader-type individuals. These communities are the basis of the information flow between the different harems, thus they can be viewed rather as alliances between harems than real communities. Overlaps between different harems, led by different leaders, are very rare. Most of the harems are connected with each other through the leader of the harem, but they are connected through a few edges between two follower-type individuals, as well. If the summarized link capacity of all leader-types is abundant, than communities, that consist only of follower-type individuals, do not survive. Follower-type individuals tend to follow only one leader-type individual, despite the fact that their followings are basically spontaneous. Therefore in the model we define a harem as the list of the follower-type individuals, who consider the choice of the given leader-type individual, including the leader as well. With this definition a harem can contain some follower-type individuals as well, who do not belong to the harem forming communities. As the network converges, the difference between the two harem definitions decreases. However, after $1000$ time steps some differences remain, and there are some individuals who are not participating in any community. The network inside a harem is very hierarchical, with one leader-type individual on the top level, and strong hierarchy exists between the harem members as well. There is also a tree-like hierarchy between harems typically with one single leader harem on the top. It is identified as the hierarchy of harem-leaders and represents the next level of organization. The whole herd keeps cohesive, since the network of individuals forms one connected cluster. The network of harems in most of the cases also remains cohesive, but if the number of leader-type individuals is relatively low, it can break up into smaller networks, as discussed later.

In analysing the model, we identified some input parameters, such as the ratio of leader-type individuals to all individuals, and the typical value of link capacities, which influence the results more sensitively. Other parameters, like group size, frequency of herding behaviour in a round and the number of considered decisions by an individual have less considerable effect. The frequency of herding behaviour affects first of all the rate of convergence to the stable state, but not the resulting network. The higher the frequency, the faster the convergence is.  

\begin{figure}[p]
\centering
\includegraphics[scale=0.44]{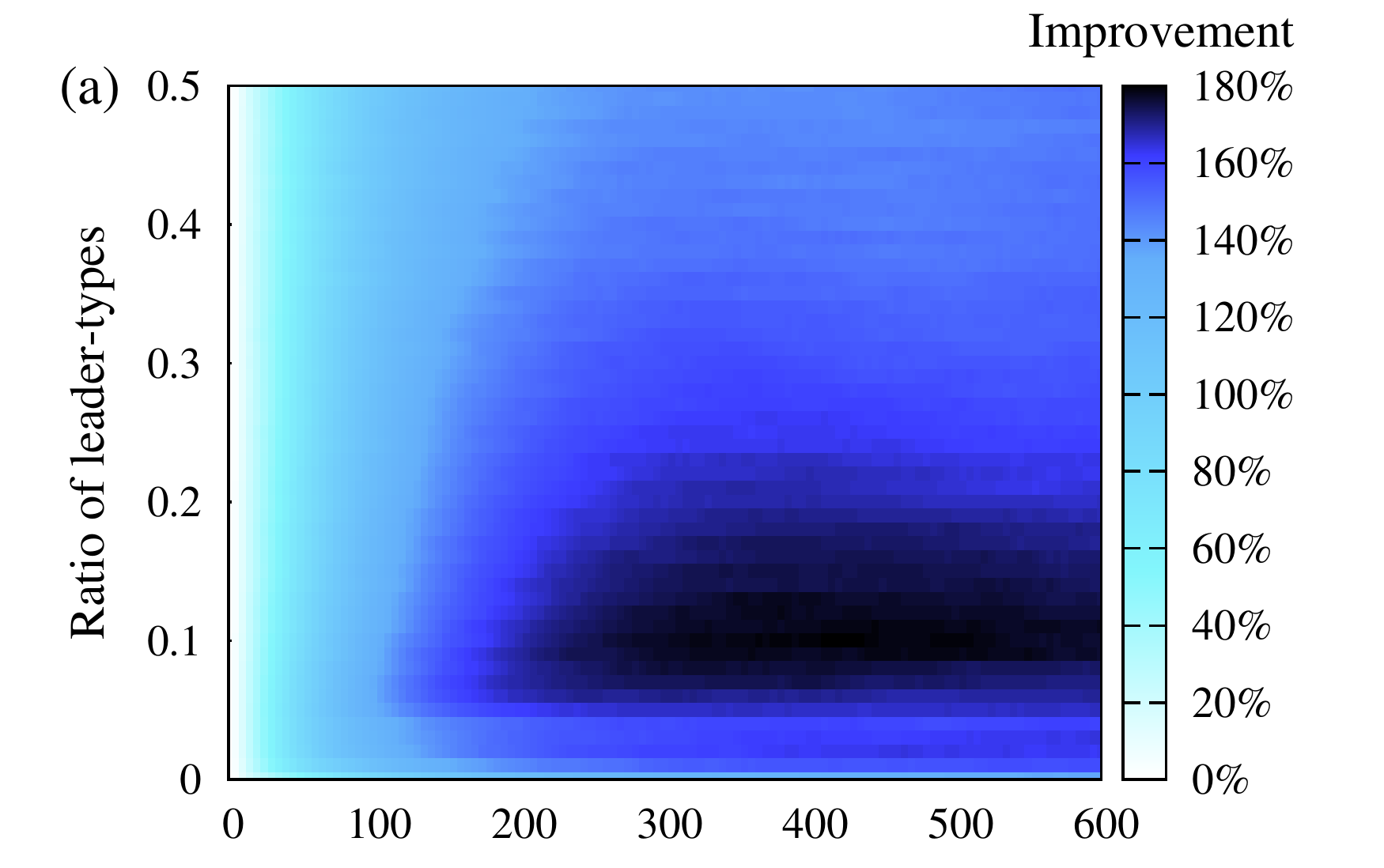}\includegraphics[scale=0.44]{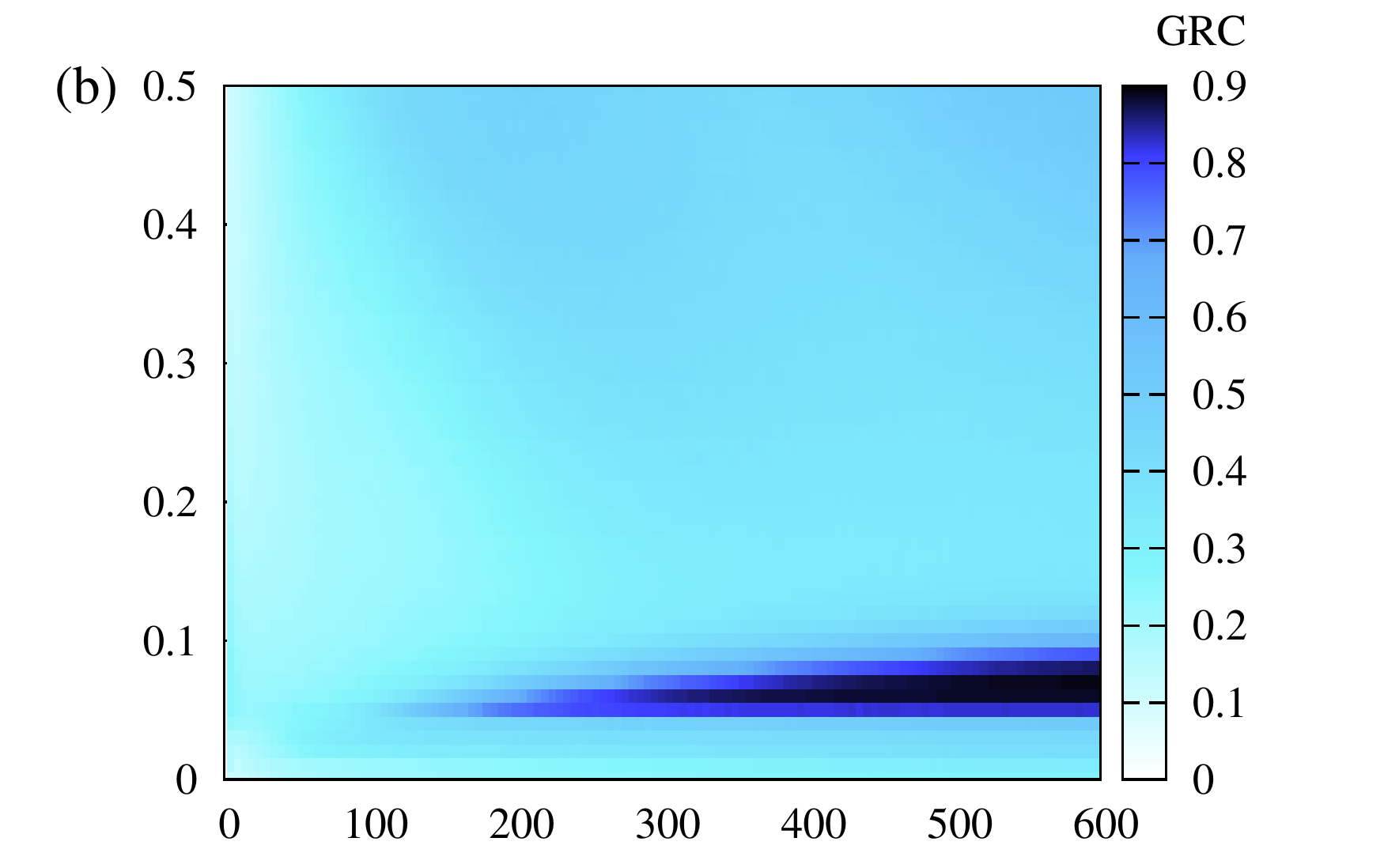}
\includegraphics[scale=0.44]{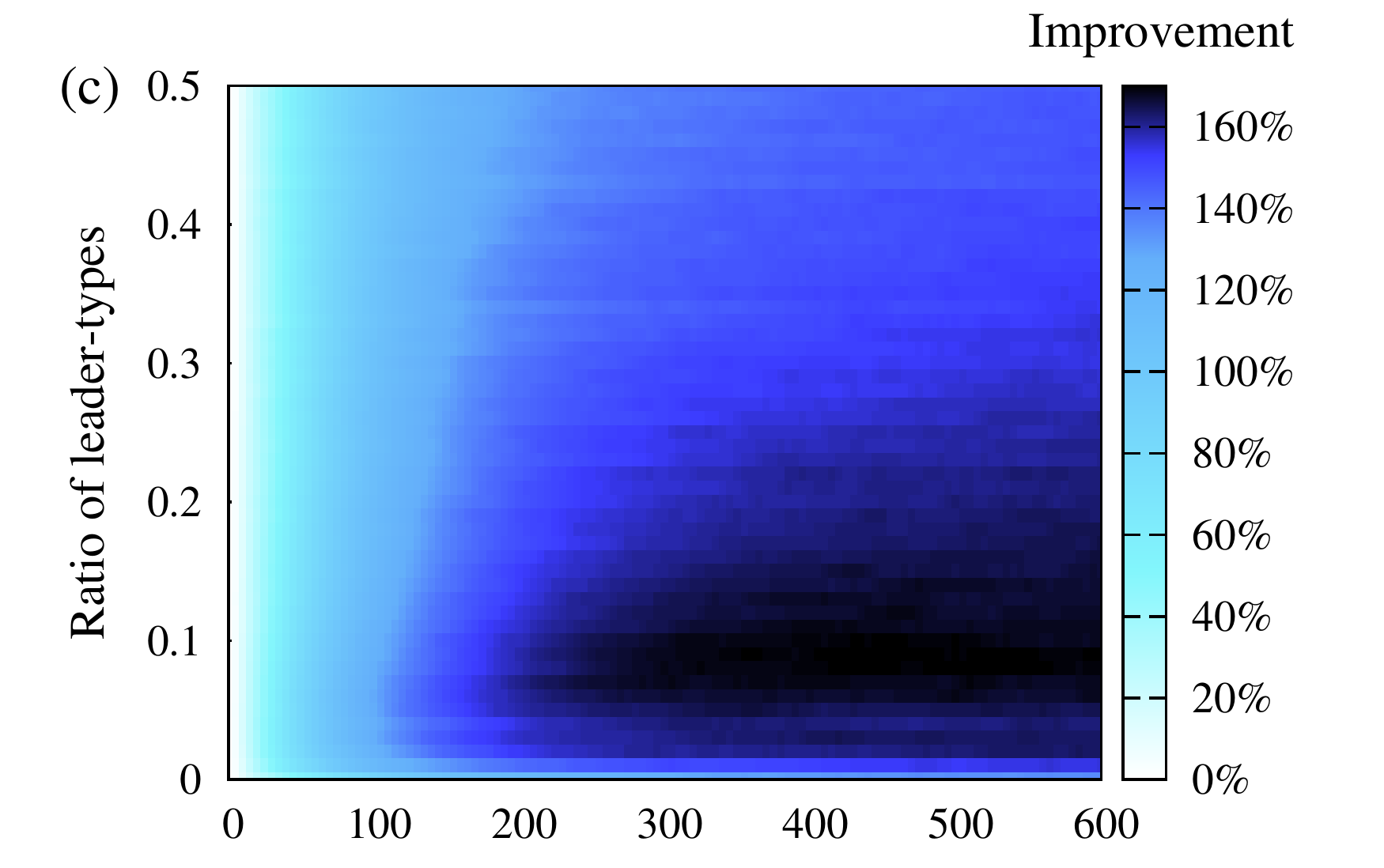}\includegraphics[scale=0.44]{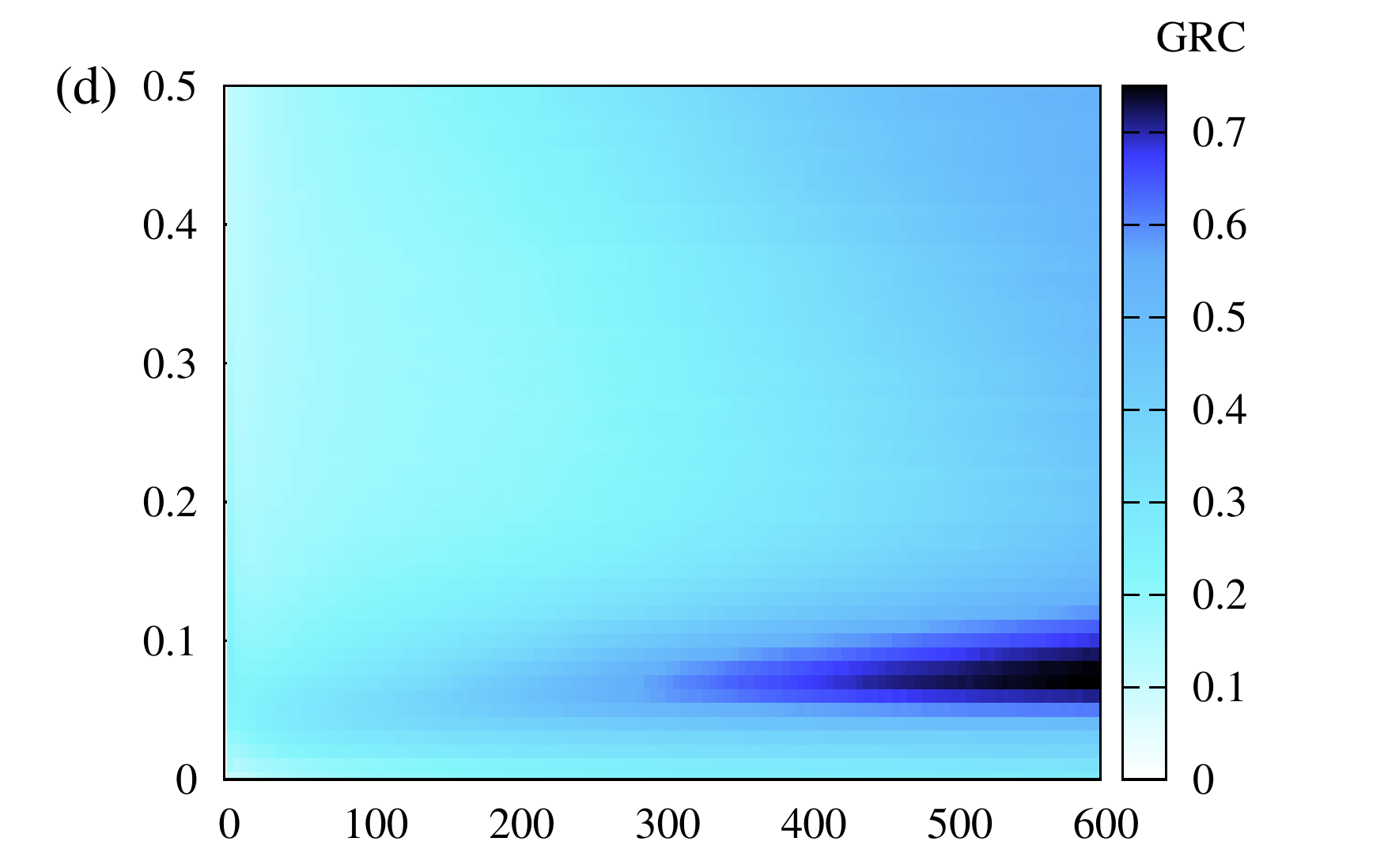}
\includegraphics[scale=0.44]{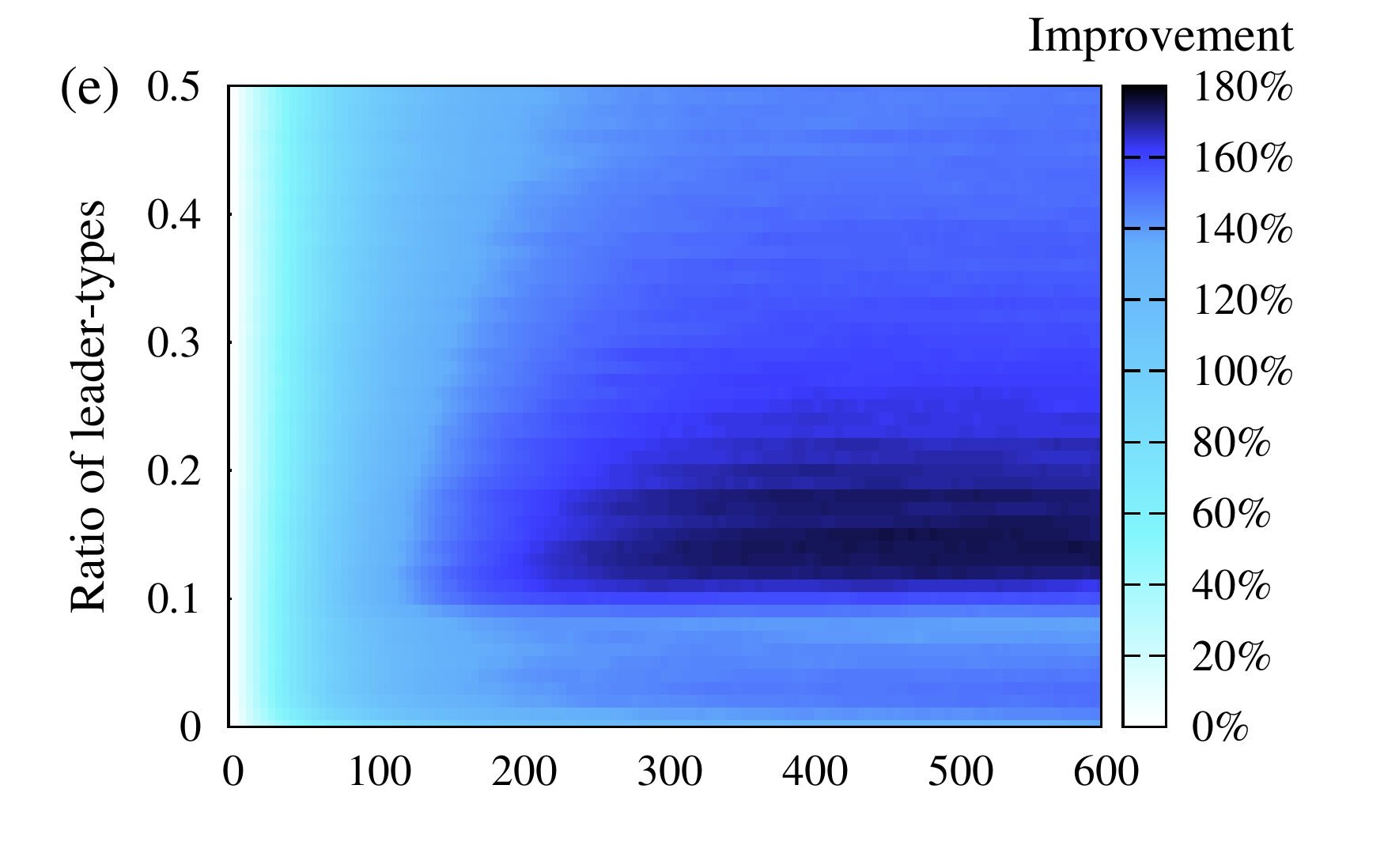}\includegraphics[scale=0.44]{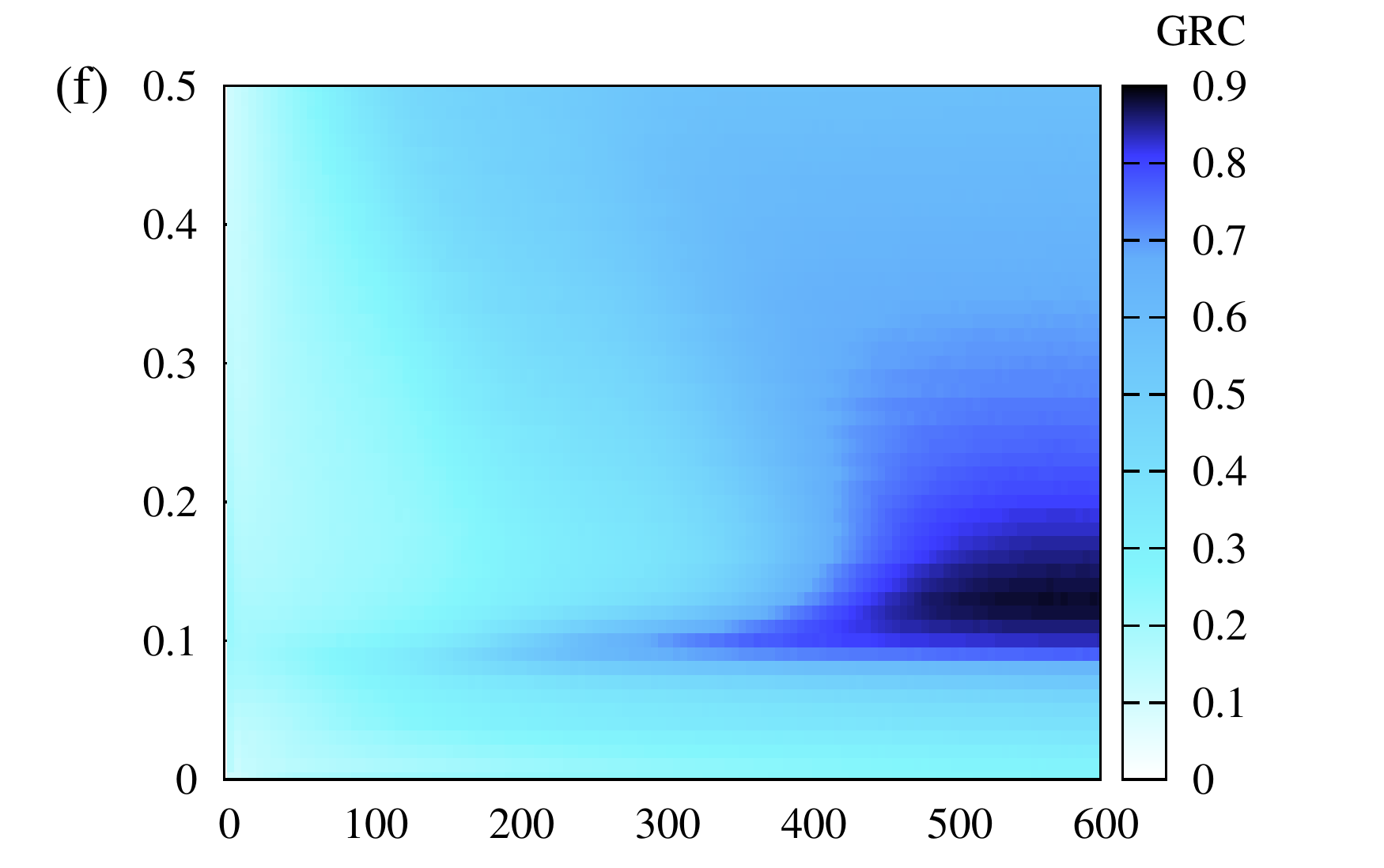}\\
\vspace{-0.2cm}\includegraphics[scale=0.44]{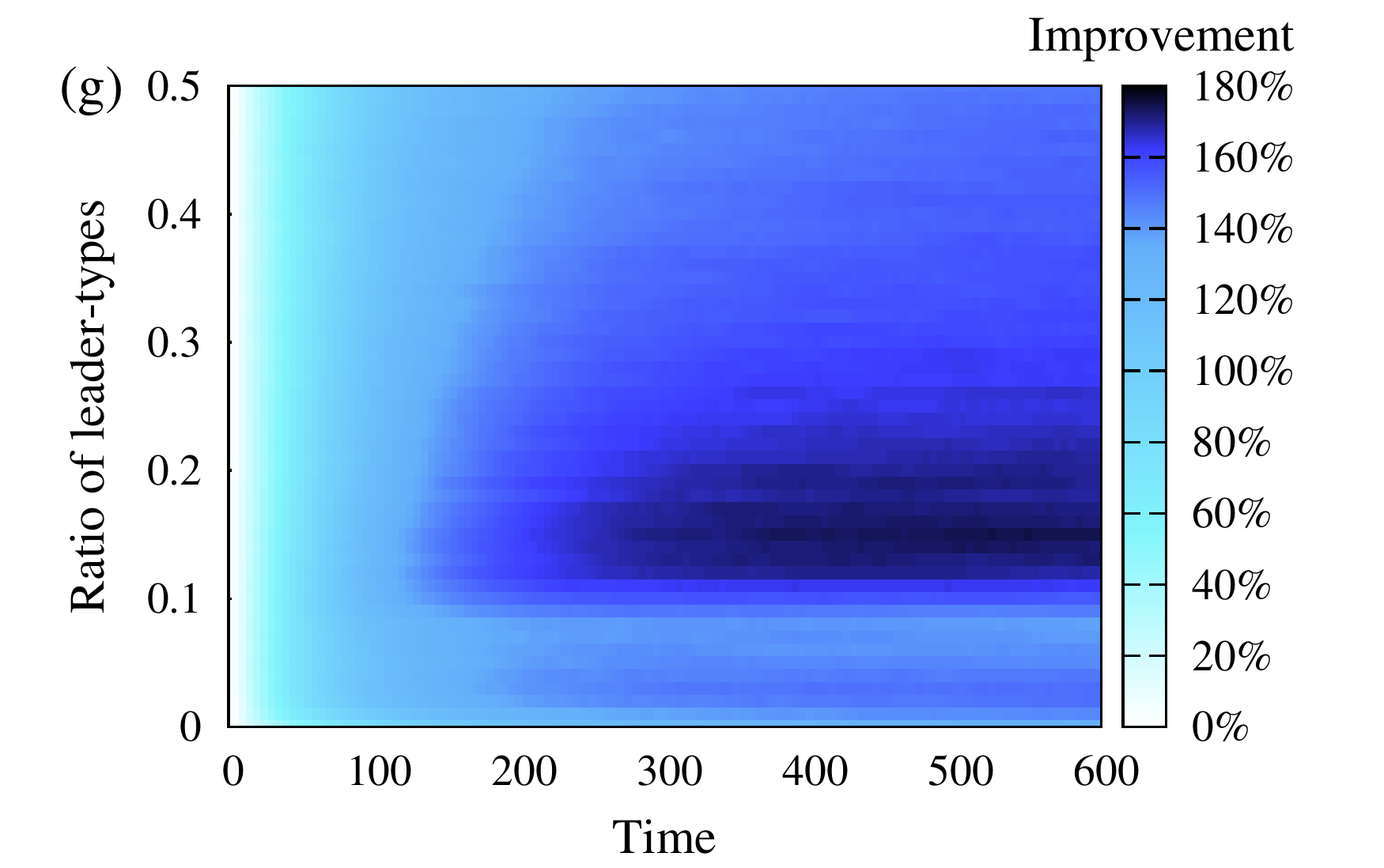}\includegraphics[scale=0.44]{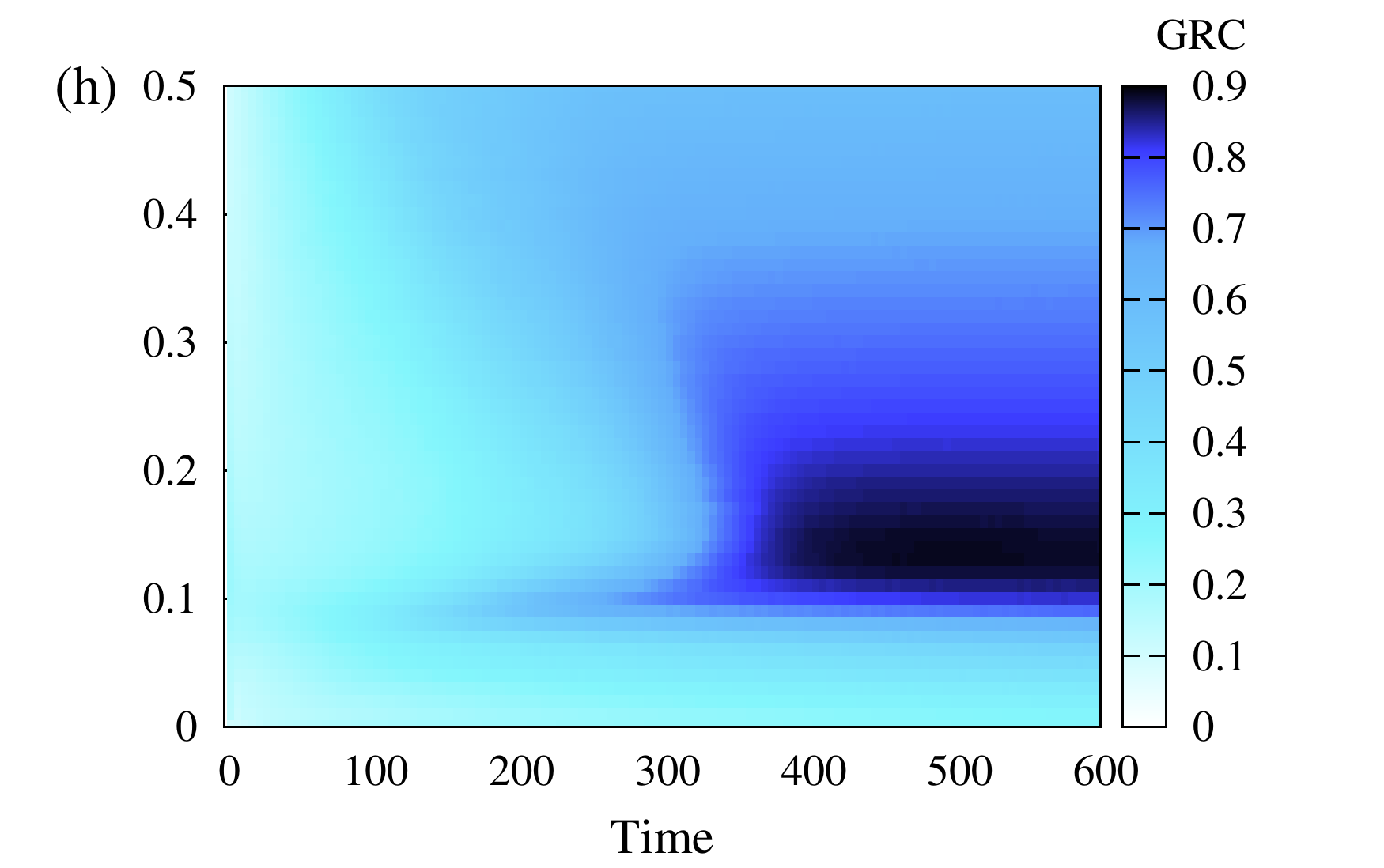}\vspace{-0.5cm}
\caption{Heatmaps of the relative performance improvement (\ref{imp}) (left column) and global reaching centrality (GRC) \cite{mones} (right column) of model networks, as the function of time and ratio of leader-type individuals to the number of all individuals. An optimal harem leader to harem member ratio can be observed, where performance and hierarchy is maximized. The location of the optimum is robust to changes in the shape of the link capacity distribution (LCD) if its average is fixed: (a)-(b) is for a Poisson LCD with $\lambda = 20$, and (c)-(d) for a lognormal LCD with $\mu=20$ mean and $\sigma=20$ standard deviation. For lower average LC values the optimal range is shifted: (e)-(f) is for a Poisson LCD with $\lambda = 10$, and (g)-(h) for a $(3,17)$ uniform LCD. The number of all individuals in the simulations is $n = 200$, their ability distribution is bounded Pareto with $0.25$ expected value and $1/\sqrt{48}$ standard deviation, each data point is averaged over $1000$ runs.}
\label{heatmaps}
\end{figure}

\subsection{Experiment and model}
\begin{figure}[H]
\includegraphics[scale=0.45]{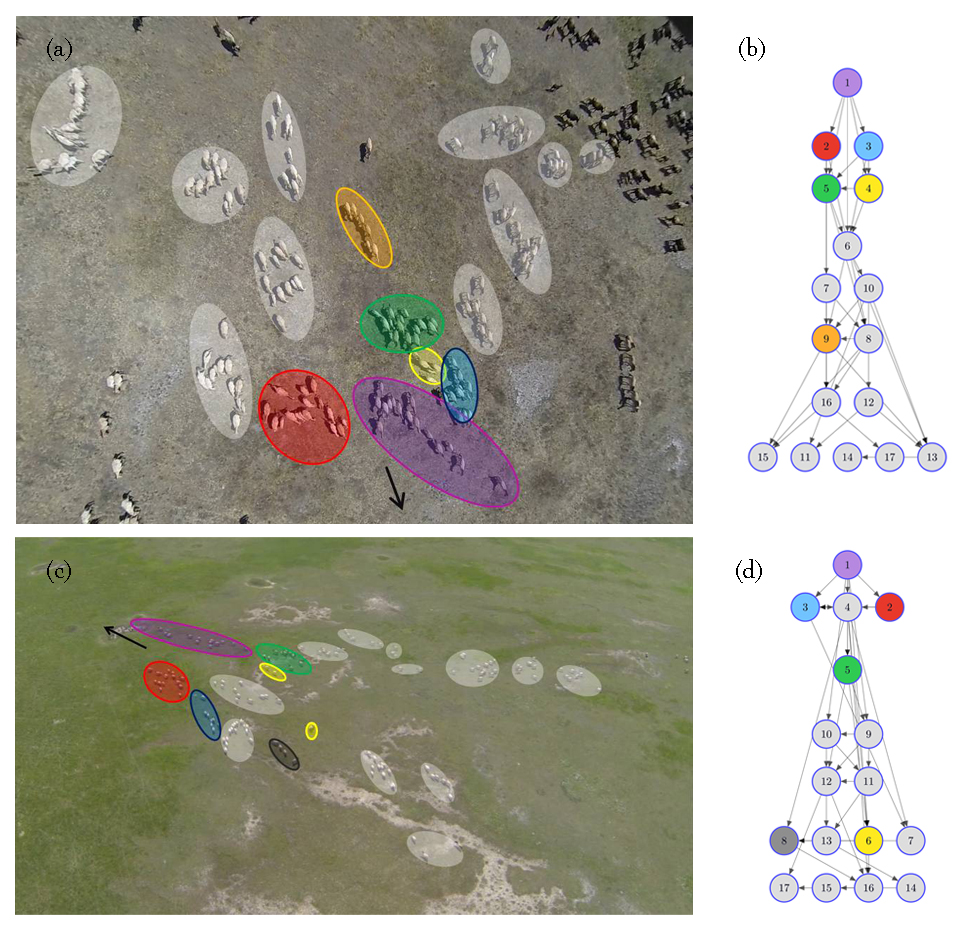}
\caption{Aerial pictures taken on 22.07.2013 (top) and on 21.05.2014 (bottom). The highlighted areas indicate the individual harems. The black arrow shows the direction of motion. The color-codes match the same stallions$'$ harems on both pictures. The identification is based on the number of adult and infant horses in the harems compared with the up-to-date catalogue of the reserve (note that these numbers can differ on the two pictures due to the elapsed time). The violet, red, blue, yellow and green harems move at the front of the V-formation of the herd, and thus are supposed to play a leader role, in both pictures. On the basis of the above figures it is reasonable to assume that - on average - the leadership roles of the harems remained about the same over the 10 months indicated above. The figures on the right show the hierarchical layout of the estimated leadership network of harems visualized with the reaching centrality method \cite{mones}.}
\label{s1}
\end{figure}

\begin{figure}[H]
\includegraphics[scale=0.45]{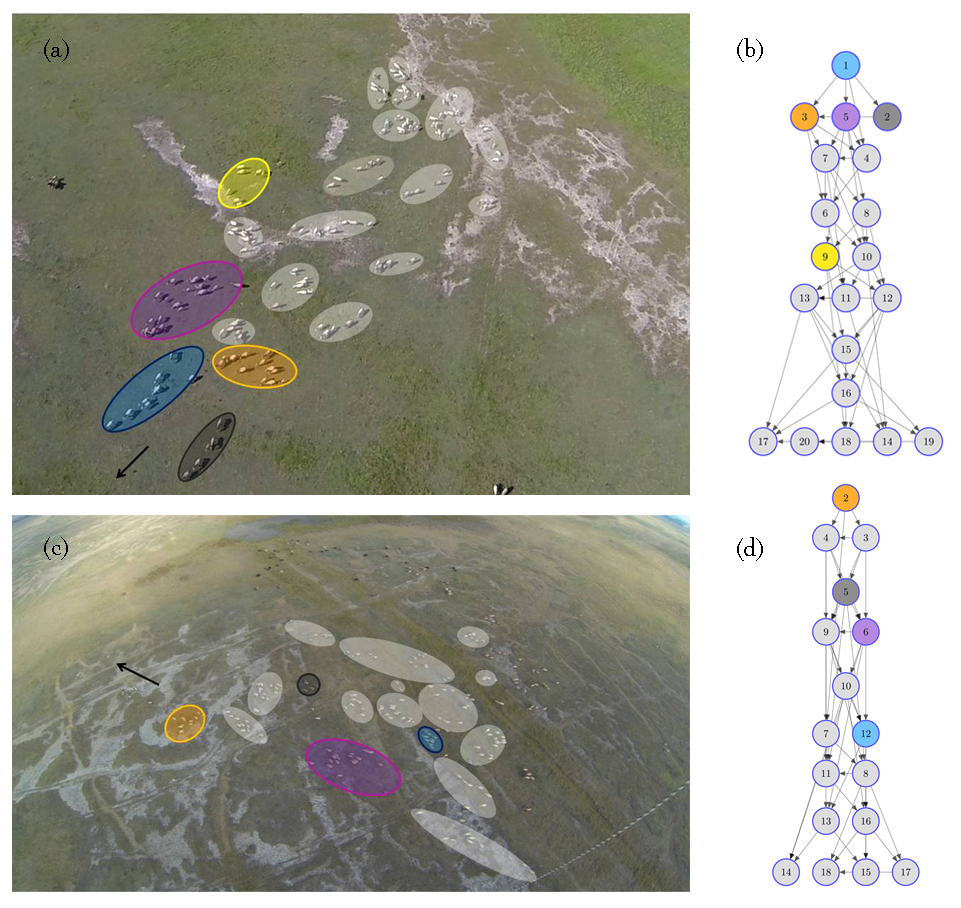}
\caption{Aerial pictures taken on 05.06.2014 (top) and 06.07.2014 (bottom). The blue and violet harems that move at the front area of the herd on 21.05.2014 (Fig. S1/c) are still in leading positions on 05.06.2014 (a), but later on 06.07.2014 (c) they can be found in the middle area. The harems marked with red and green in Fig. S1 could not be identified here. The yellow harem that could be identified only in (a), moves further in the herd as time passes (Fig. S1/a, S1/c and S2/a). The orange and black harems that move at the front both on 05.06.2014 (a) and 06.07.2014 (c), are middle positioned on Fig. S1/a and S1/c, respectively. In comparison with the situation on 21.05.2014 (Fig. S1/c) the leadership roles have somewhat changed. We conjecture that the changes which can be detected on Fig. S1/c, S2/a and S2/c and refer to the supposed leadership structure between 21.05.2014 and 06.07.2014 are likely to be due to the breeding season that takes place in the late spring.}
\label{s2}
\end{figure}

Carrying out simulations with similar input parameters as those observed in the wild horses, very similar networks can result both qualitatively and quantitatively on average. Based on the aerial images (e.g. Fig.~\ref{horses}) and the network definition in Sect.~\ref{observation}, the hierarchy of harems can be established (Fig.~\ref{comparison} (a) and Fig.~\ref{s1}, \ref{s2}). The number of identified harems in Fig.~\ref{horses} and thus the number of nodes is $17$. The harems contain roughly a total number of $150$ horses. A typical harem network of the model is shown in Fig.~\ref{comparison} (b). The number of all individuals is $n=150$, and the number of leader-type individuals is $m=17$ in the simulation, in order to fit the experimental parameters. Both the experimental and model networks are visualized with reaching centrality method \cite{mones}. At first glance one can see the similar pyramid-like layout of the networks and the common features, such as the presence of one single leader, several nodes in higher layers, and many at the bottom layer. There are many layers that indicates the varying roles, and the edges can connect distant layers. The global reaching centrality (GRC) values of the model networks are very close to those of the experimental networks, on average, quantifying a similar level of hierarchy in the two cases. Particularly, in Fig.~\ref{comparison} GRC takes the values $0.65$ and $0.67$ for the experimental and model networks, respectively. However, the number of edges is less in the model ($26$) than in the experiment ($36$) in this case. The enlarged areas in the right display the internal structure in some harems.

\subsection{Optimal harem leader to harem member ratio}

In order to find the main features of the model networks we carried out simulations with a range of input parameters and calculated network properties (group performance and hierarchical measures) in every case by averaging over $1000$ runs. The investigation shows that the presence of the leader-type individuals results in an increment in relative performance improvement (\ref{imp}) from about $135\%$ to as high as $170\%$. In addition, the resulting network is more hierarchical according to all the three measures studied (fraction of noncyclic edges, global reaching centrality and fraction of forward arcs). For example GRC, can increase from $0.25$ up to $0.9$ for particular input parameters. The despotic approach of our model removes many of the non-efficient cycles. As a consequence, the decisions of top ranking individuals spread more effectively to lower ranking individuals, thereby improving overall success.

The ratio of leader-type individuals to all individuals (abbreviated as leader ratio) plays an important role in determining the quality of the resulting network. The simulations indicate that there is an optimum in the value of this parameter (Fig.~\ref{heatmaps}), where performance and hierarchy is maximized. For example, for Poisson link capacity distribution (LCD) with $\lambda=20$ the optimal leader ratio lies around $1:10$, where the relative performance improvement reaches $180\%$ after several hundred steps, and networks producing performance above $170\%$ lie in the range from $1:20$ to $1:5$ leader ratio, while outside this region performance does not exceed $160\%$ (Fig.~\ref{heatmaps} (a)). The GRC has an optimal region as well, where its value reaches $0.9$, in contrast to $0.4$ outside this region (Fig.~\ref{heatmaps} (b)). However, the optimal region of the GRC is narrower than the one of the performance, it lies between $1:20$ and $1:10$. The overlap between them, and thus the optimum from the point of view of performance and degree of hierarchy together, is between $1:20$ and $1:10$. 

The location of the optimal range is robust to changes in the shape of the link capacity distribution, if its average value is fixed. We examine uniform, delta, Poisson and lognormal link capacity distributions with averages ranging from 5 to 20. Fig.~\ref{heatmaps} (c)-(d) shows results for lognormal LCD with $\mu=20$ expected value and $\sigma=20$ standard deviation. In comparison with Fig.~\ref{heatmaps} (a)-(b), the optimal range remains the same, and the performance is maximized around $1:10$ leader ratio, similarly as for the underlying Poisson LCD with $\mu=20$ average. For lower average link capacity values a minimum of performance appears below the optimal area, see Fig.~\ref{heatmaps} (e) and (g) for a Poisson and uniform LCD with $\mu=10$ mean. If the number of leader-type individuals is decreased to the extent that $m\left\langle c_i \right\rangle < n-m$, i.e. the sum of all link capacities becomes less than the number of follower-type individuals, then link capacities start to limit the free formation of harems. If this threshold is reached the network of harems starts to break apart, since the edges connecting leader-type individuals also start to split up, giving rise to separated harems, and this can cause a decrease in overall performance. Decreasing $m$ still further can slightly increase performance, this can occur because the size of the separated communities increases. The optimal range is shifted upwards, probably because an optimal network structure can emerge when the total number of link capacities is abundant. For an underlying LCD with $\mu=10$ average, the optimal leader to harem-member ratio is around $1:8$. Comparison of simulations with different average link capacities leads to the conclusion that the optimal region lies around $m\approx1.5 n/\left\langle c_i \right\rangle$.  

The observed maximal harem size in wild horses is around $20$ in our experiment and the average size is $9$. Therefore the theoretical average link capacity (corresponding to the case of the wild horses) can be between $10$ and $20$, giving rise to an optimal harem-leader to harem-member ratio of roughly between $1:8$ and $1:10$, from the point of view of the common success. It is very interesting that the $1:9$ empirical ratio observed in wild horses is close to this model result.

\subsection{Harem size distribution}

\begin{figure}[bt]
\vspace{-6cm}\hspace{-1.25cm}\includegraphics[scale=0.45]{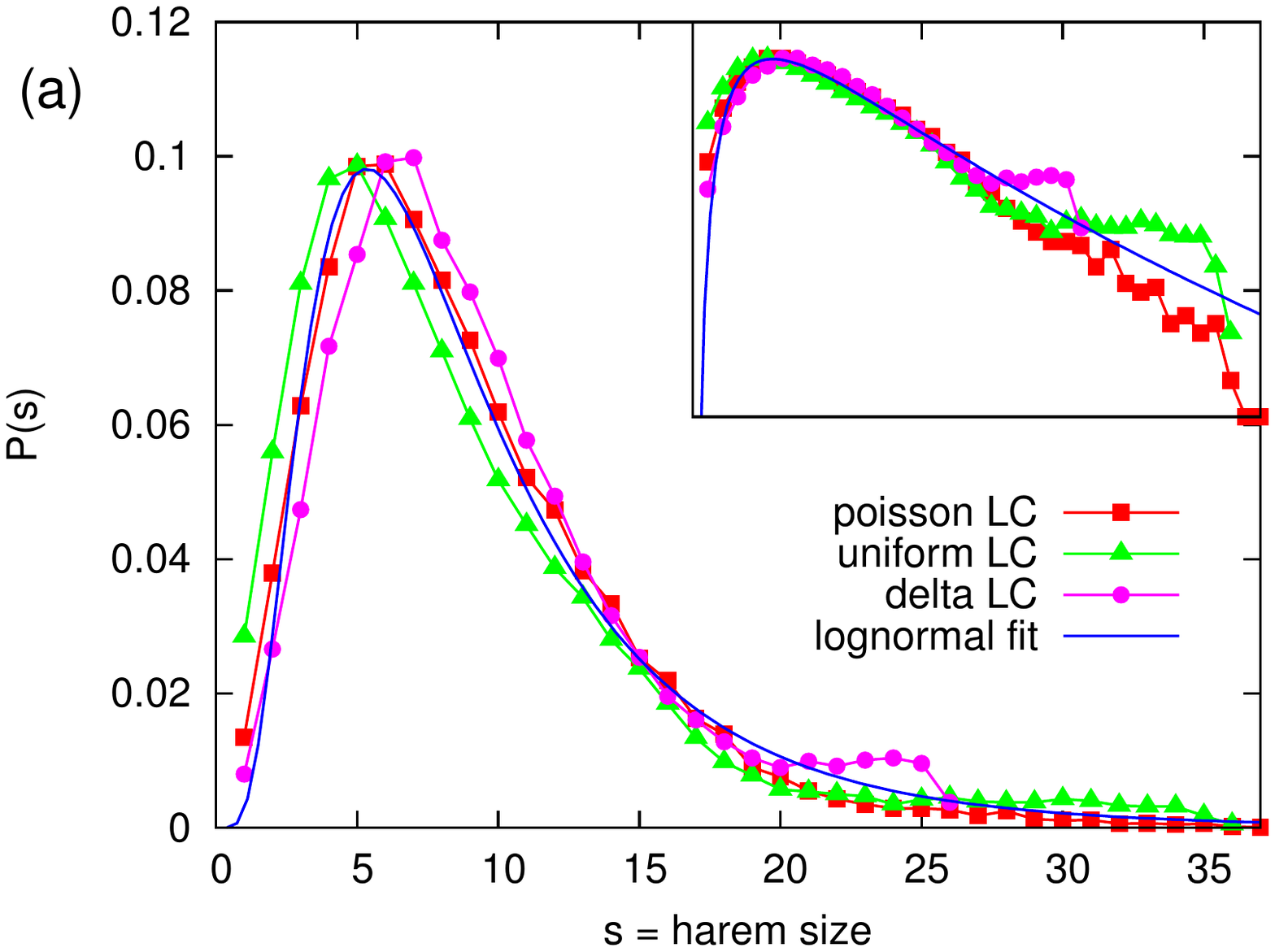}\hspace{-1.5cm}\includegraphics[scale=0.45]{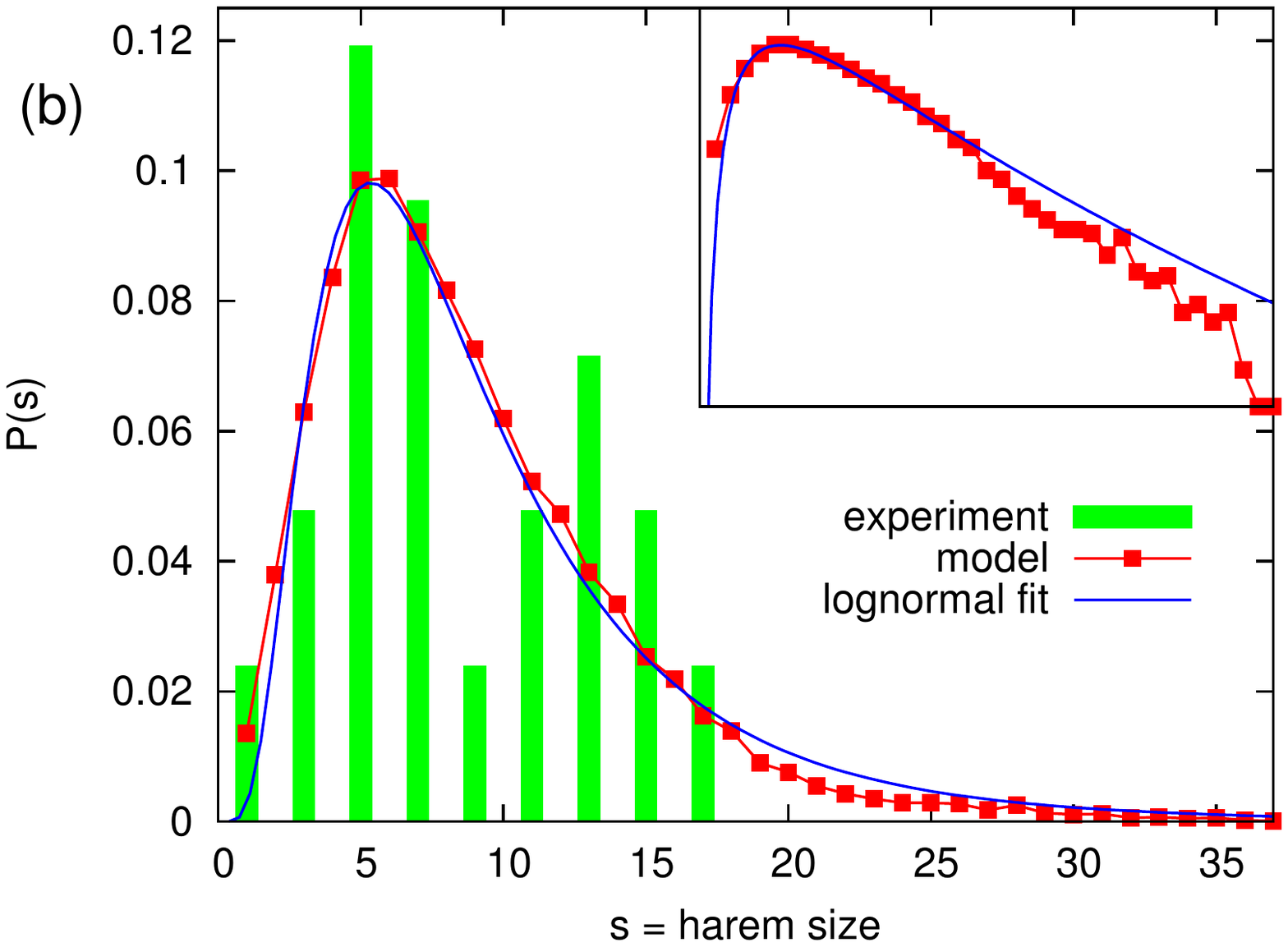}
\vspace{-1.5cm}\\
\caption{Harem size distribution. (a) Model distributions tend to follow a lognormal distribution, provided the total number of link capacities is large enough, and it does not limitate the harem formation. It does not depend on the shape of the link capacity distribution: harem size distributions with underlying delta($25$), uniform($15,35$) and Poisson($\lambda=25$) LCDs are shown on the plot. A lognormal distribution is fitted to the case of the Poisson LCD with $\mu=9.61$ mean and $\sigma=6.65$ standard deviation. (b) The experimental harem size distribution is plotted together with the model distribution (with an underlying Poisson LCD with $\lambda = 25$) and the lognormal fitted to the model. Using the Kolmogorov-Smirnov test the fitted lognormal is accepted as the theoretical distribution of the experimental sample at $p=0.97$ significance level. The experimental distribution is based on data from $m = 21$ wild horse harems including $n = 188$ individuals. The number of all individuals in the simulation is $n = 200$, and $m = 25$ of which are leader-type individuals. Each data point is averaged over $1000$ runs with harem sizes measured at the $t = 1000$ time step. Semi-log plots are shown in the top right corner.}
\label{clusters}
\end{figure}

Since our model aims to simulate a group of groups, it is natural to ask what kind of cluster size distributions (in our case, harem size distributions) characterize the resulting network. The harem sizes are defined as the number of follower-type individuals following the given harem leader, and the harem leader is also counted in the size. When the network, and thus the harem sizes, can be considered as converged, we build a histogram and investigate it for different LC distributions. In order to have a roughly optimal performing network in all cases, when simulating with LC distributions with average values ranging from $10$ to $25$, the leader ratio is set to $1:8$. We find that the distribution of harem sizes is an asymmetric heavy-tailed distribution, and can be well fitted by a lognormal (Fig.~\ref{clusters} (a)). Similarly to the optimal leader ratio, it seems to be independent of the shape of the link capacity distribution. It is very similar for uniform, delta and Poisson distributed link capacities (Fig.~\ref{clusters} (a)). However, this only holds if the predefined LCD of the individuals does not limitate the harem formation. Again, if $m\left\langle c_i \right\rangle < n-m$, leader-type individuals fill up all their links and a trivial harem size distribution emerges, with a shape determined by the link capacity distribution. As the total number of links is increased, the harem size distribution approaches a lognormal. 

The comparison of the model with the experiment also shows encouraging agreement. The experimental harem size distribution is based on the data from $m = 21$ wild horse harems including $n=188$ individuals, and its histogram is shown in Fig.~\ref{clusters} (b). Despite the considerable error due to the small sample size, an accordance can be proposed with the lognormal harem size distribution coming from the model. We assume the null hypothesis, that the theoretical distribution of the observed harem sizes is the lognormal fitted to the model results for a Poisson ($\lambda=25$) LCD, with fitting parameters $\mu=9.61$ mean and $\sigma=6.65$ standard deviation. Using the Kolmogorov-Smirnov test it can be accepted at $p=0.97$ significance level that the experimental sample comes from this theoretical distribution. 
 
\section{Discussion}
\label{discussion}

The endangered status of the Przewalski horses calls for more profound studies of the overall behaviour of this species \cite{houpt}, and indeed, their collective movements have attracted interest as well \cite{bourjade}. However, the different patterns of their collective motion, and the leadership structure behind it, has not been investigated in detail yet. A proper analysis of the individual horse tracks would clarify our estimated picture, and would deepen the understanding of the organization and cohesion of harems. Although, our estimation can give a valuable insight into the leadership hierarchy of the horse harems, since it reflects the possible leader-follower relationships provided that the horses use visual perception when following others. 

Our model simulates the emergence of a leadership hierarchy in a three-level herd starting from unfamiliar individuals, and results in a stable state representing a consistent leadership network of a natural group. It is based on the individuals copying the behaviour of some of the herd mates when making individual decisions. Collective decision emerges from these individual ones. The specific feature of our present approach is that in line with the spontaneously developing leader-follower relationships between the individuals, dense sub-groups are forming inside the herd that are loosely connected with each other, similarly to the social structure of the wild horses. In the converged network the hierarchy of the sub-groups is similar to the hierarchy of the individuals inside a sub-group, typically including a single leader. Thus, our model accounts for three basic levels of social organizations in a decision-making context. 

Modeling shows that the harem-leader to harem-member ratio observed in the herd of Przewalski horses corresponds to an optimal network in the sense that the overall success of the group is maximized. In addition, hierarchy is maximized with this ratio as well, suggesting that the degree of hierarchy in a group correlates with common success. The observed distribution of harem sizes is consistent with the lognormal distribution obtained when applying our model. The above results are not too model specific. The model behaves similarly for different values of the main parameters - like group size, the typical number of other group mates whose behaviour a member considers, before making its decision, and the distribution of the maximal possible harem size a leader is able to keep together. A better statistics for the harem sizes of horses and a generalization to other species would improve the understanding of the underlying processes.

Finally, we find that a modularly structured leadership hierarchy is more beneficial than a non-modular one when making a collective decision. The spreading of advantageous decisions of higher ability individuals is more facile through mid-level leaders, thus approving the overall success of the herd. This observation suggests that multilevel societies, beside kin relationships and the formation of breeding groups, may be the results of alternative optimization factors, as well.

\begin{acknowledgements}
We are grateful to Kristin Brabender and the Directorate of Hortob\'agy National Park for providing us their data as well as officially authorizing the carrying out of our observations and to Waltraut Zimmermann, K\"olner Zoo for cooperation. We thank G\'abor V\'as\'arhelyi and Gerg\H o Somorjai for helping our observations with flying robots. This research was partially supported by the EU ERC COLLMOT project (grant no. 227878).
\end{acknowledgements}

\beginsupplement
\section*{Supplementary Information}
\large{Additional figures of the Przewalski horse herd (Hortob\'agy National Park, Hungary) during collective movements}

\end{document}